**Title: Mid-infrared Observations of the Nucleus of Comet P/2016 BA$_{14}$ (PANSTARRS)**

Authors: Takafumi Ootsubo[1], Hideyo Kawakita[2,3], Yoshiharu Shinnaka[2]

[1] Astronomy Data Center, National Astronomical Observatory of Japan, 2-21-1 Osawa, Mitaka, Tokyo 181-8588, Japan

[2] Koyama Astronomical Observatory, Kyoto Sangyo University, Motoyama, Kamigamo, Kita-ku, Kyoto 603-8555, Japan

[3] Department of Astrophysics and Atmospheric Sciences, Faculty of Science, Kyoto Sangyo University, Motoyama, Kamigamo, Kita-ku, Kyoto 603-8555, Japan

Abstract:

We present mid-infrared observations of comet P/2016 BA$_{14}$ (PANSTARRS), which were obtained on UT 2016 March 21.3 at heliocentric and geocentric distances of 1.012 au and 0.026 au, respectively, approximately 30 hours before its closest approach to Earth (0.024 au) on UT 2016 March 22.6. Low-resolution ($\lambda/\Delta\lambda \sim 250$) spectroscopic observations in the N-band and imaging observations with four narrow-band filters (centered at 8.8, 12.4, 17.7 and 18.8 μm) in the N- and Q-bands were obtained using the Cooled Mid-Infrared Camera and Spectrometer (COMICS) mounted on the 8.2-m Subaru telescope atop Maunakea, Hawaii. The observed spatial profiles of P/2016 BA$_{14}$ at different wavelengths are consistent with a point-spread function. Owing to the close approach of the comet to the Earth, the observed thermal emission from the comet is dominated by the thermal emission from its nucleus rather than its dust coma. The observed spectral energy distribution of the nucleus at mid-infrared wavelengths is consistent with a Planck function at temperature $T \sim 350$ K, with the effective diameter of P/2016 BA$_{14}$ estimated as ~0.8 km (by assuming an emissivity of 0.97). The normalized emissivity spectrum of the comet exhibits absorption-like features that are not reproduced by the anhydrous minerals typically found in cometary dust coma, such as olivine and pyroxene. Instead, the spectral features suggest the presence of large grains of phyllosilicate minerals and organic materials. Thus, our observations indicate that an inactive small body covered with these processed materials is a possible end state of comets.

1. INTRODUCTION

Comets are icy small bodies passing through the solar system that contain potentially invaluable information concerning the origins of the materials forming the solar system. In periodic comets that have orbited the Sun for many times with progressively smaller perihelion distances, the nucleus exhibits signs of significant physical and chemical evolution (Mumma & Charnley 2011). Such evolutionary effects are most prominent at the nucleus surface. To distinguish the pristine properties of cometary nuclei from these evolutionary effects, clarifying the influence of solar heating on cometary nuclei is imperative. Thus, conducting observational and theoretical studies on evolved comets is essential.

The evolutionary track of extremely weakly active comets is considered to culminate in their dormancy or extinction. There are several plausible end states for comets including dormancy, extinction (plausibly connected to primitive asteroids such as P- or D-type asteroids), fragmentation (e.g., C/1999 S4 (LINEAR) and C/2012 S1 (ISON); Farnham et al. 2001; Sekanina & Kracht 2014), collision with a planet or the Sun (e.g., D/1993 F2 (Shoemaker–Levy 9; Levy 1998)), and removal from the inner solar system due to gravitational perturbation by giant planets (Fernandez 2005). Because cometary dormancy and extinction are considered gradual processes (Li et al. 2017), comparing the physical properties of cometary nuclei with possible dormant/exhausted comet candidates, such as D-type asteroids, is important for verifying the proposed links between comets and primitive asteroids (Campins et al. 2018).

Herein, we report on the mid-infrared photometric and spectroscopic observations of comet P/2016 BA$_{14}$ (PANSTARRS), whose gas and dust production rates, according to imaging and spectroscopic observations at optical wavelengths (Li et al. 2017; Hyland et al. 2019), were revealed to be extremely low even near its perihelion passage at 1.012 au from the Sun. As discussed by Li et al. (2017), comet P/2016 BA$_{14}$ is likely approaching its end state. Note that an active comet (252P/LINEAR) has similar orbit to comet P/2016 BA$_{14}$ (see Table 1). These comets have almost identical Tisserand invariant parameters with respect to Jupiter, which indicates that these comets have undergone a strong gravitational interaction with Jupiter. These characteristics make plausible the hypothesis that the two comets are fragments of the same initial body. In contrast to comet P/2016 BA$_{14}$, comet 252P/LINEAR was active during its perihelion passage in the 2016 apparition and its chemical composition suggest a rather typical composition in short-period comets surveyed at infrared wavelengths (Paganini et al. 2019) and in comets surveyed at optical wavelengths (Li et al. 2017; Schleicher 2008). Thus, in contrast to its current depleted state, comet P/2016 BA$_{14}$ might originally have been similarly rich in volatile materials as comet 252P/LINEAR if one assumes that the two objects have a common origin.

The close approach of comet P/2016 BA$_{14}$ to the Earth, 0.024 au on UT 2016 March 22.6, enabled observations of the inner coma and nucleus of the comet (Li et al. 2017; Naidu et al. 2016). Radar observations revealed that P/2016 BA$_{14}$ has a nucleus greater than 1 km in diameter. Its absolute magnitude of 19.5 and a diameter of at least 1 km correspond to an optical albedo of < 3% (Naidu et al. 2016). Furthermore, Li et al. (2017) also estimated the nucleus size of P/2016 BA$_{14}$ as ~1 km by assuming that its total visible brightness was dominated by the nucleus signal throughout its perihelion passage and that the geometric albedo of the nucleus surface was 4%. These estimates are consistent with the estimate based on NEOWISE data (~700 m, according to a private communication from J. Bauer 2017 in Li et al. 2017). Our mid-infrared photometric and spectroscopic observations were performed at night approximately 30 hours before the closest approach of the comet to the Earth in 2016, with the mid-infrared properties of the nucleus of P/2016 BA$_{14}$ discussed herein.

Table 1: Orbital elements of P/2016 BA$_{14}$ and 252P/LINEAR.

|  | P/2016 BA$_{14}$ [a] | 252P/LINEAR [b] |
|---|---|---|
| $T$ [UT] | 2016 March 15.51740 | 2016 March 15.27847 |
| $q$ [au] | 1.0085744 | 0.9960732 |
| $e$ | 0.6663164 | 0.6736753 |
| $a$ [au] | 3.0225476 | 3.0523992 |
| *Peri.* [deg.] | 351.89525 | 343.29119 |
| *Node.* [deg.] | 180.53526 | 190.98106 |
| *Incl.* [deg.] | 18.92006 | 10.40467 |
| $T_J$ [c] | 2.82 | 2.82 |

[a] Epoch 2016 April 2.0 TT = JDT 2457480.5 (from JPL/HORIZONS)

[b] Epoch 2016 January 13.0 TT = JDT 2457400.5 (from JPL/HORIZONS)

[c] Tisserand invariant parameter with respect to Jupiter.

## 2. OBSERVATIONS

Mid-infrared imaging and spectroscopic observations of comet P/2016 BA$_{14}$ (PANSTARRS) were conducted by using the 8.2-m Subaru Telescope with the Cooled Mid-infrared Camera and Spectrometer (COMICS) (Kataza et al. 2000; Okamoto et al. 2003) on UT 2016 March 21.3, when the comet was at heliocentric and geocentric distances of 1.012 au and 0.026 au, respectively. The closest approach of P/2016 BA$_{14}$ to the Earth (0.024 au) occurred on UT 2016 March 22.6, ~30 hours after our observations. Imaging observations were performed using narrow-band filters (centered at $\lambda$ = 8.8, 12.4, 17.7 and 18.8 μm, with $\Delta\lambda$ = 0.8, 1.2, 0.9, and 0.9 μm, respectively) in the N- and Q-bands, while spectroscopic observations were obtained in the N-band with a spectral resolving power ($R = \lambda/\Delta\lambda$) of ~250, corresponding to a slit width of 0.33 arcsec (the used slit length was 40 arcsec). For the cancellation of the sky background radiation in the mid-infrared observations, secondary-mirror-chopping was used at a frequency of 0.43–0.45 Hz with an amplitude of 15 arcsec, which are common values for both spectroscopic and imaging observations. Table 2 lists the observational conditions of the comet and a standard star.

Flux calibration was performed by observing the spectro-photometric standard star HD72094 (tet Cnc) characterized by Cohen et al. (1999), while wavelength calibration was performed via comparison with the sky emission lines. We used IRAF[1] software and the special tools provided by the COMICS instrument team[2] for the data reduction. Photometry was performed with an aperture radius of 2.60 arcsec. To achieve an adequate signal-to-noise ratio (SNR) for the extracted spectra, they were extracted with a slit region of 0.33 arcsec (slit width) by 3.14 arcsec (along the slit direction).

---

[1] IRAF is distributed by the National Optical Astronomy Observatory, which is operated by the Association of Universities for Research in Astronomy (AURA) under cooperative agreement with the National Science Foundation.

[2] https://subarutelescope.org/Observing/DataReduction/index.html

Table 2: Observing conditions.

| UT Date | UT Time | Object | Integration time (s) | Observing mode | Airmass |
|---|---|---|---|---|---|
| 2016 March 21 | 08:38 | P/2016 BA14 | 10.8 | Imging/N8.8 | 1.12 |
| | 08:35 | | 10.6 | Imging/N12.4 | 1.11 |
| | 08:51 | | 30.5 | Imging/Q17.7 | 1.15 |
| | 08:46 | | 30.5 | Imging/Q18.8 | 1.14 |
| | 08:42 | | 30.7 | Spectrosc./NLspc[a] | 1.13 |
| 2016 March 21 | 09:07 | HD72094 | 15.4 | Imaging/N8.8 | 1.16 |
| | 09:04 | | 15.4 | Imaging/N12.4 | 1.15 |
| | 08:54 | | 60.1 | Imaging/Q17.7 | 1.13 |
| | 08:59 | | 60.1 | Imaging/Q18.8 | 1.14 |
| | 09:11 | | 30.7 | Spectrosc./NLspc[a] | 1.17 |

(a) Low-resolution spectroscopic mode ($R \sim 250$) in the N-band.

3. RESULTS

Figure 1 shows the spectral energy distribution of P/2016 BA$_{14}$ based on our photometric measurements. The observations are consistent with the characteristic blackbody radiation observed for many other comets (Ootsubo et al. (2020); Shinnaka et al. (2018); and references therein). The observations are matching a Planck function at temperature $T$~350 K (the best-fit temperature is 343 ± 10 K in consideration of flux errors), as shown in Figure 1. The equilibrium temperature of a blackbody at $r$ [au] from the Sun is expressed as ~280 $r^{-0.5}$ [K] while the observed equilibrium temperature of comet is slightly higher than this. This excess relative to the equilibrium temperature of a blackbody (the so-called superheat) is observed frequently for cometary coma grains because the dominant grains in the coma have sub-μm dimensions and are smaller than the mid-infrared wavelengths of the emitted radiation (i.e., the mid-infrared radiation cooling of these small grains is inefficient). However, the spatial profiles of P/2016 BA$_{14}$ and a standard star (as a point source) are almost identical, with no extensions observed for P/2016 BA$_{14}$ when it was compared to a point source, as shown in Figure 2. Thus, the thermal radiation from cometary coma grains was not collected effectively by the slit. In such case, a scaling factor for the Planck function plotted in Figure 1 is proportional to a product of surface area of the cometary nucleus and its emissivity in mid-infrared wavelengths (i.e., by assuming the nucleus as a gray-body). The contribution of the nucleus signal and the estimated effective radius of the nucleus are discussed in the next section.

Figure 3 shows the N-band low-resolution (R ~ 250) spectrum of P/2016 BA$_{14}$, to which we applied a running average technique using a 7-pixels window to remove noise spikes. Additionally, we applied the slit loss correction by adjusting the spectrum to fit the photometric data points at 8.8 and 12.4 μm. The mid-infrared spectrum of the comet is relatively smooth, with shallow absorption-like features in this wavelength region. In contrast to typical active comets, the spectrum recorded for P/2016 BA$_{14}$ exhibited no prominent 10-μm excess emission feature.

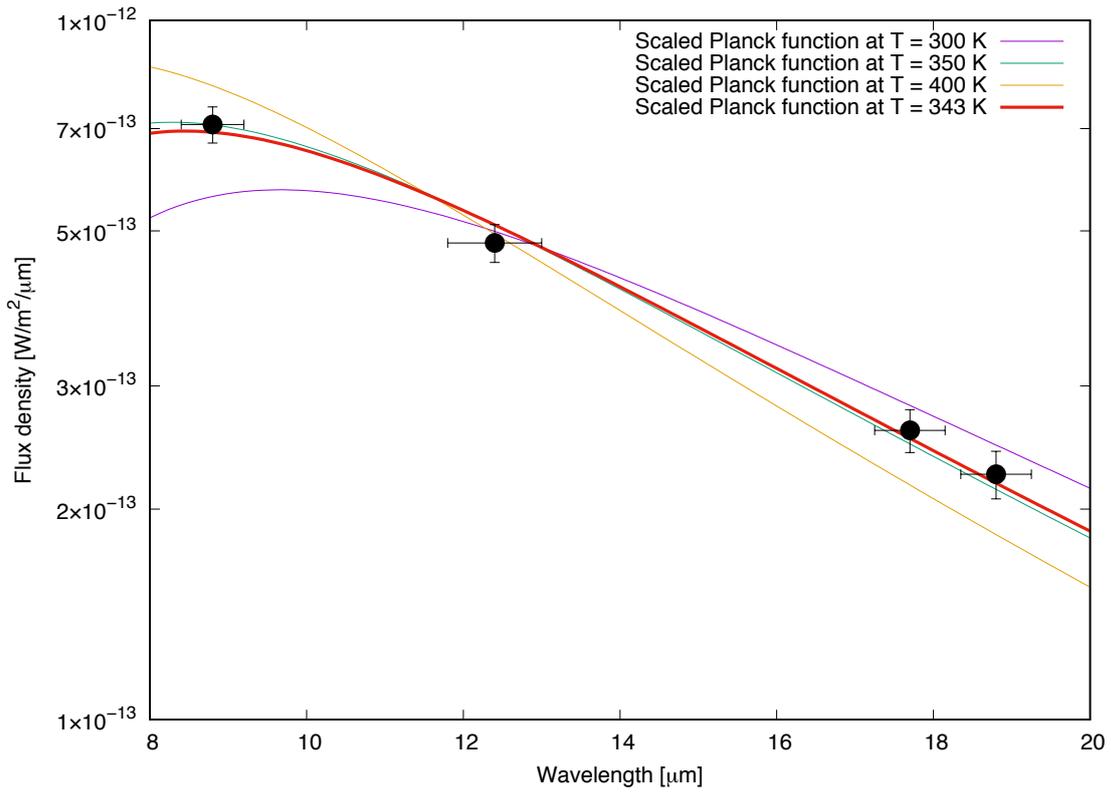

Figure 1: Photometric results (Mid-IR spectral energy distribution) of P/2016 BA$_{14}$. Vertical bars on the data points (filled black circles) are photometric errors and horizontal bars denote band widths of the used filters. The photometric observations are consistent with a blackbody radiation at temperature $T$~350 K.

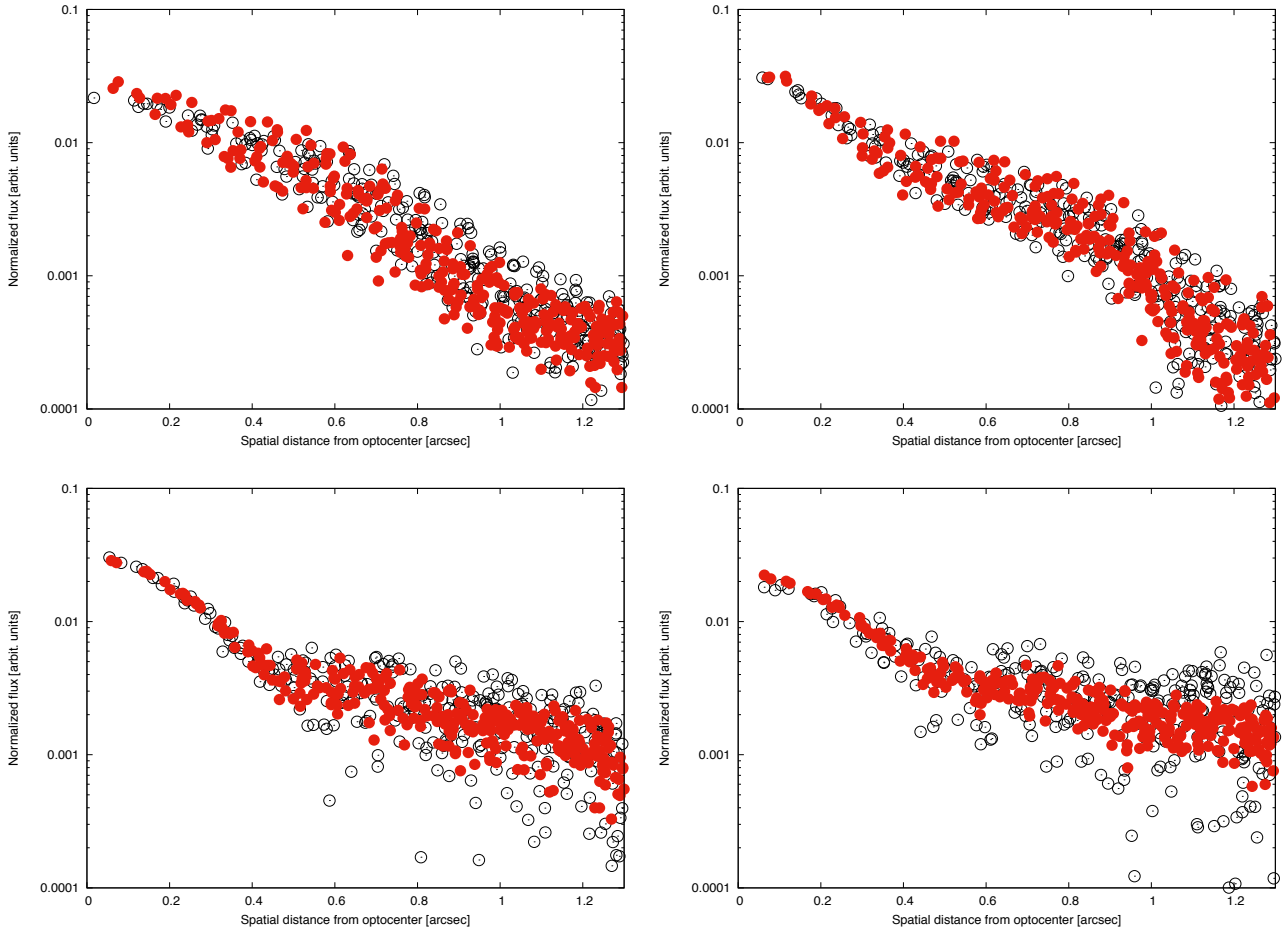

Figure 2: Spatial profiles of P/2016 BA$_{14}$ (filled circles) compared with a standard star HD72094 (open circles) at different wavelengths: (clockwise from top-left) 8.8, 12.4, 18.8, and 17.7 μm. Vertical axes are in relative flux normalized by total flux within aperture (a radius of 1".3). The spatial profiles of P/2016 BA$_{14}$ are consistent with a point-spread-function (i.e., the characteristic spatial profile of a point source, HD72094).

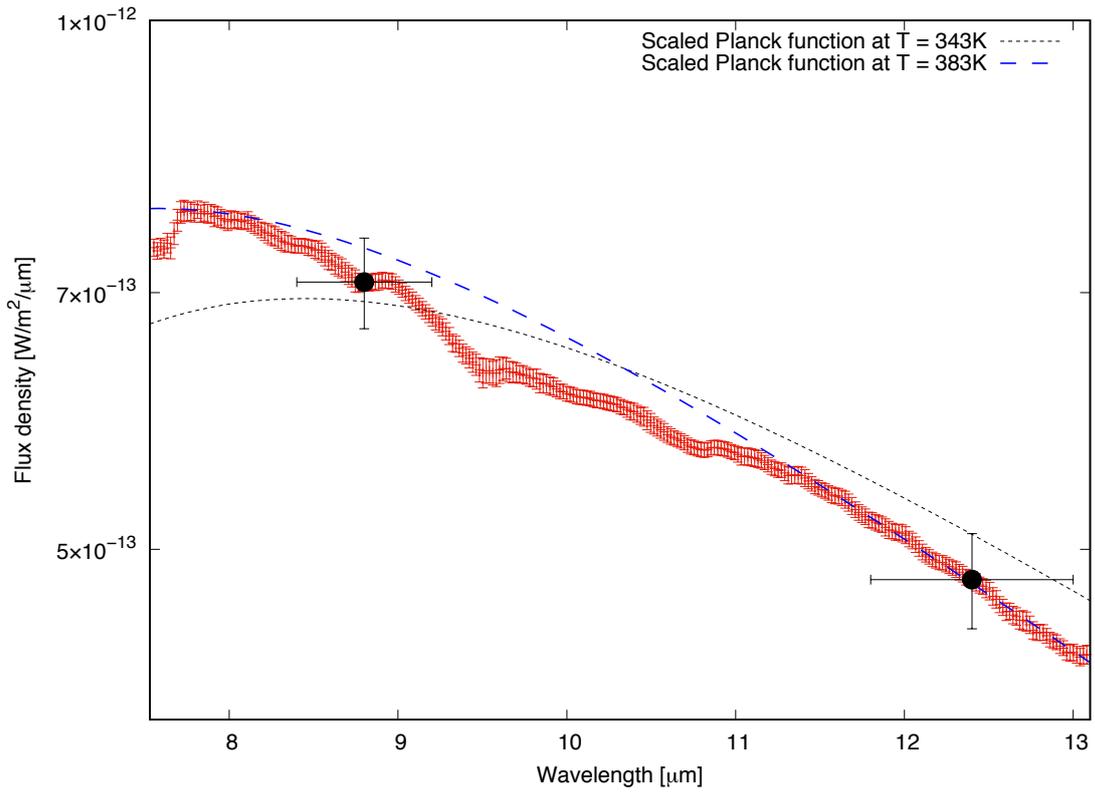

Figure 3: N-band low resolution spectrum of P/2016 BA$_{14}$ (red data; SNR improved by taking the running average using a 7-pixels window) together with the photometric data points at 8.8 and 12.4 μm (black circles). Error-bars of the N-band spectrum correspond to relative error levels. Scaled Planck functions at temperature $T$ = 343 and 383 K are also plotted (dotted and dashed lines, respectively).

## 4. DISCUSSION AND CONCLUSIONS

As shown in Figure 2, the spatial profiles observed for comet P/2016 BA$_{14}$ are consistent with those of a photometric standard star (a point source). The thermal emission from the nucleus of P/2016 BA$_{14}$ appears dominated by contribution from mid-infrared photons, and the coma signal was negligible. In the thermal wavelength region, the dominant grains in the coma (in the sub-μm diameter range) are smaller than the wavelength of the emitted radiation (i.e., the smaller grains are not efficient infrared emitters). In contrast, the cometary nucleus (which size is much larger than the wavelength) can emit thermal infrared photons effectively. Therefore, for an extremely weakly active comet, the signal from the nucleus dominates the observed infrared signal, and the contribution of dust grains in coma is negligible (e.g., Kelly et al. 2017). The dominance of the signal from the nucleus for spectral energy distribution and the N-band low-resolution spectrum of P/2016 BA$_{14}$ (Figures 1 and 3) is attributed to the very low gas and dust production rates of P/2016 BA$_{14}$ around its perihelion passage (Li et al. 2017; Hyland et al. 2019).

Figure 1 shows that the spectral energy distribution of P/2016 BA$_{14}$ is consistent with radiation from a blackbody (or gray-body) object at 343 ± 10 K. As discussed earlier, the thermal radiation observed in P/2016 BA$_{14}$ was dominated by the nucleus signal, and we should compare the observed equilibrium temperature with the surface temperatures of cometary nuclei. Suttle et al. (2020) compiled a list of the measured comet surface temperatures for five comets: 1P/Halley, 9P/Tempel 1, 19P/Borrelly, 67P/Churyumov-Gerasimenko, and 103P/Hartley 2 (Emerich et al. 1986; Soderblom et al. 2004; Groussin et al. 2007, 2013; Tosi et al. 2019) and derived a formula for the surface diurnal temperature; $T = 339.95\ r^{-0.438}$ [K] for a heliocentric distance of $r$ [au] from 0.8 to 3.4 au. Our measurement for P/2016 BA$_{14}$ at 1.012 au is consistent with the measurements reported for these comets. Furthermore, based on our photometric measurements and assuming an emissivity of 0.97 (corresponding to an assumed optical albedo of the nucleus surface of 0.03), the effective diameter of P/2016 BA$_{14}$ is estimated to be 0.8±0.2 km for a surface temperature $T = 343$ K, that is derived with 8.8–18.8 μm. This is consistent with previous estimates, which predict the nucleus size to be ~1 km or larger (Naidu et al. 2016; Li et al. 2017). Note that a scaled Planck function at temperature $T = 343$ K cannot reproduce the observed N-band spectrum well as shown in Figure 3. This is probably caused by a lack of photometric data points to be fitted in wavelength region shorter than ~8 μm (a peak of the Planck function might be at a shorter wavelength than the wavelengths we observed). The observed N-band spectrum can be fitted by a scaled Planck function at temperature $T = 383$ K (Figure 3). Although we simply assume a uniform surface temperature to derive an effective diameter, the surface temperature might not be uniform. If we consider the uncertainty of surface temperature, the effective diameter of the nucleus of P/2016 BA$_{14}$ is estimated to be in the range of 0.5–1.2 km.

Figure 4 shows the emissivity spectrum of P/2016 BA$_{14}$ (we used a scaled Planck function at temperature $T = 383$ K to normalize the observed mid-infrared spectrum). The emissivity spectrum could be represented by a linear combination of six Gaussian components (whose parameters are listed in Table 3), as shown in Figure 4. The most prominent features are centered at 7.59, 9.50, and 10.6 μm. In general, larger mineral grains on the surface of cometary nuclei are considered responsible for absorption-like features in thermal emissivity spectra, while excess emission features are representative of μm-sized small grains (Emery et al. 2006). As shown in Figure 4, the spectrum of P/2016 BA$_{14}$ differs from the thermal emissivity spectrum corresponding to the cometary nucleus of comet 10P/Tempel 2, which was inactive at a further distance from the Sun (Kelley et al. 2017). Conversely, the thermal emissivity spectrum of the nucleus of 10P/Tempel 2 showed a emission-like excess feature at ~10 μm, in contrast to the absorption-like features observed in P/2016 BA$_{14}$. The thermal emissivity spectrum of 10P/Tempel 2 more closely resembles those of D-type asteroids, which some consider to be exhausted or dormant comets (Figure 4 also shows the thermal emissivity spectrum of the D-type asteroid 624 Hektor; Emery et al. 2006). Note that the emissivity spectra exhibiting an excess feature at ~10 μm are similar to those of cometary coma, for which thermal emission is produced by fine grains in the coma. This suggests that the nuclei surfaces of 10P/Tempel 2 and asteroids such as 624 Hektor are covered by highly porous grains (Kelley et al. 2017; Vernazza et al. 2012), which are probably composed of the anhydrous Mg-rich silicate minerals reported for the grains in cometary coma (Shinnaka et al. 2018; Bardyn et al. 2017). Because the effective diameter of 10P/Tempel 2 was estimated to be ~9 km (Kelley et al. 2017) and thus considerably larger than P/2016 BA$_{14}$ (~1 km), the larger surface gravity of the nucleus of 10P/Tempel 2 might retain a greater number of smaller grains on its surface, thereby producing the 10-μm excess emission feature.

It could be argued that the absorption-like features in the normalized thermal emissivity spectrum of P/2016 BA$_{14}$ are caused by larger silicate grains because surfaces covered by these grains (> 100 μm) give rise to thermal emissivity spectra exhibiting absorption-like features (Emery et al. 2006; Hamilton 2010). In the case of crystalline silicates, some sub-peaks are usually found in their thermal emissivity spectra (see Figure 10 in Hamilton 2010). Figure 5 shows a comparison with a thermal emissivity spectrum estimated using a reference reflectance spectrum for Mg-rich natural olivine (Fo92) grains ranging from 75 to 250 μm (from the ECOSTRESS spectral library; https://speclib.jpl.nasa.gov/; Baldridge et al. 2009), together with the thermal emissivity spectra representing a surface covered with large porous grains comprising Mg-rich olivine and pyroxene in both crystalline and amorphous forms (Fabian et al. 2001; Jaeger et al. 1998; Dorschner et al. 1995). The latter were based on a combination of Hapke theory (Hapke 2012) and Mie theory (Bohren & Huffman 1983) and assumed a grain diameter of 1 mm and a volume porosity of 30%. Both peak positions and the relative strengths of the sub-peaks don't match between the observed spectrum and the spectrum for Mg-rich natural olivine in laboratory. Although the peak wavelengths of the features depend on the Mg:Fe ratio in olivine (longer wavelengths correspond to a larger Fe content), the change in peak strength ratios are not drastic (Hamilton 2010). The calculated emissivity spectra for crystalline olivine and pyroxene cannot explain the features in the emissivity spectrum of P/2016 BA$_{14}$, while the absorption features calculated for amorphous olivine and pyroxene are significantly broader than those observed for P/2016 BA$_{14}$. Finally, we note that laboratory spectra and optical constants used for calculating modeled spectra shown in Figure 5 were considered to be measured at temperatures (i.e., room temperatures; ~300 K), different from the temperature of cometary surface (~350 K). We assumed that emissivity peaks don't change significantly in the range of 300–400 K although the emissivity peaks (or absorption peaks) in thermal infrared wavelength region are known to have temperature dependence (e.g., Chihara et al. 2001).

By comparing the normalized thermal emissivity spectrum of P/2016 $BA_{14}$ with the thermal emissivity spectra measured for various minerals in the laboratory (Lane & Bishop 2019), the shape and position of the observed ~10 μm feature observed for P/2016 $BA_{14}$ are more similar to those of phyllosilicates (such as chlorite, clinochlore, and serpentine) rather than anhydrous silicates. Although the thermal emissivity spectrum of cometary dust coma is usually modeled with anhydrous silicates (see Shinnaka et al. 2018 and references therein), Lisse et al. (2006, 2007) considered phyllosilicates when modeling the dust grains of comets 9P/Tempel 1 and C/1995 O1 (Hale-Bopp), although phyllosilicates were minor components of the overall cometary grain composition.

We also compared the spectrum of P/2016 $BA_{14}$ to that of the (entirely rock-based) chondrite meteorite Orgueil (CI), as shown in Figure 4. The mineralogy of Orgueil is dominated by fine-grained phyllosilicates (various forms of the serpentine and the clay mineral saponite), which were probably formed during hydrothermal alteration of the parent body that might be a comet (Gounelle & Zolensky 2014). The shape of the absorption-like feature at ~10 μm is similar for the thermal emissivity spectra of P/2016 BA14 and the meteorite Orgueil. However, their minima occur at different wavelengths; the minimum of the absorption-like feature for P/2016 $BA_{14}$ is at 9.50 μm, whereas the minimum for Orgueil occurs at ~10.0 μm. This difference could be the result of structural and phasal differences between the silicate minerals (Suttle et al. 2017; Che & Glotch 2012).

Suttle et al. (2017) reported on the thermal infrared reflectance spectra of fine-grained micrometeorites and classified them into different groups (groups 1–5) corresponding to varying degrees of thermal processing, which were most likely caused during their entry into the telluric atmosphere. Group 1 corresponds to dehydrated and unheated chondritic materials (phyllosilicates dominate the mineralogy of a hydrated chondritic matrix), and members of this group show thermal reflectance spectra similar to those of carbonaceous chondrites (CI, CM, and C2 ungrouped), with a single peak whose maxima (i.e., minima in thermal emissivity spectra) lies at approximately 10 μm. Alternatively, Groups 2 and 3 correspond to dehydroxylate mineraloids (heated to ~600–1000 K), which are characterized by a relatively sharp and dominant peak at 9.0–9.5 μm. Group 4 comprises partially annealed, mixed dehydroxylates and olivine (heated to ~700–1100 K) while Group 5 contains completely crystalized olivine (heated to ~1000–1500 K). As shown in Figure 6, the prominent absorption-like feature observed in P/2016 $BA_{14}$ (whose minimum is at 9.50 μm) most closely resembles a sample in the Group 3 (although the position of ~9.5 μm feature in the observed spectrum matches the Groups 3 and 4, the Group 4 displays ~11.5 μm feature which is not seen in the observed spectrum). This feature is associated with the stretching of Si-O bonds in the residual, isolated silica tetrahedra in dehydroxylated phyllosilicates (Suttle et al. 2017).

This suggests that the nucleus surface of P/2016 BA$_{14}$ was likely heated to ~600 K at least (Suttle et al. 2020). However, because the perihelion distance of P/2016 BA$_{14}$ in its current orbit is 1.012 au from the Sun, the expected surface temperature of the cometary nucleus at its perihelion passage is much lower, ~350 K (Figure 3 in Suttle et al. 2020), and cannot reach ~600 K. Therefore, this comet might have had smaller perihelion distances in the past, with a perihelion distance of ~0.27 au corresponding to a temperature of ~600 K. The current orbit of the comet has an aphelion distance of 5.25 au close to the Jovian orbit. Thus, a gravitational perturbation by Jupiter might be responsible for any past alteration in the perihelion distance of the comet. Alternatively, the original body of comet P/2016 BA$_{14}$ might have contained abundant dehydroxylated phyllosilicates grains in the interior of the nucleus. If so, the mineralogy of dust grains in the coma of comet 252P/LINEAR (which is considered as a comet pair of P/2016 BA$_{14}$, and in contrast to P/2016 BA$_{14}$, it was active; Li et al. 2017) should also be dominated by dehydroxylated phyllosilicates. This remains to be investigated in a future study.

Thus, we conclude that the prominent absorption-like feature peaked at 9.50 μm recorded in the normalized thermal emissivity spectrum of P/2016 BA$_{14}$ is associated with dehydroxylated phyllosilicates on the nucleus surface. Unfortunately, we could not identify materials as carriers of other features at 7.59, 8.35, and 8.75 μm. The carriers of these features might not be minerals and most plausible carriers are organic materials on the nucleus surface (Poch et al. 2020; Raponi et al. 2020). Table 4 lists the candidates of hydrocarbons for those features (Coates 2000; Evans et al. 2005; Draine & Li 2007). Our observations indicate that one possible end state of comet may be an inactive small body covered with coarse grains of phyllosilicate minerals combined with organic materials.

Table 3: Gaussian components fitted with observed emissivity spectrum of P/2016 BA$_{14}$.

| I.D. # | λ [μm] | FWHM [μm] | Peak depth |
|---|---|---|---|
| 1 | 7.59±0.01 | 0.16±0.01 | 0.055±0.002 |
| 2 | 8.35±0.01 | 0.43±0.02 | 0.021±0.001 |
| 3 | 8.75±0.01 | 0.32±0.02 | 0.036±0.001 |
| 4 | 9.50±0.01 | 0.77±0.03 | 0.086±0.001 |
| 5 | 10.0±0.01 | 0.38±0.04 | 0.029±0.004 |
| 6 | 10.6±0.02 | 0.90±0.03 | 0.052±0.001 |

Table 4: Candidates of carrier for unidentified features. [a,b,c]

| I.D. # | λ [μm] | Candidate | Organic species |
|---|---|---|---|
| 1 | 7.59±0.01 | Aromatic C-C stretch | Aromatic ring (aryl) group |
|   |   | Aromatic (primary, secondary, or tertiary) amine, C-N stretch | Amine and amino compound group |
|   |   | Primary or secondary, O-H in-plane bend | Alcohol and hydroxy compound group |
|   |   | Phenol or tertiary alcohol, O-H bend | Alcohol and hydroxy compound group |
| 2 | 8.35±0.01 | Tertiary amine, C-N stretch | Amine and amino compound group |
|   |   | Aromatic C-H in-plane bend | Aromatic ring (aryl) group |
| 3 | 8.75±0.01 | Secondary amine, C-N stretch | Amine and amino compound group |
|   |   | Aromatic C-H in-plane bend | Aromatic ring (aryl) group |

[a] Coates (2000).
[b] Draine & Li (2007).
[c] Evans et al. (2005).

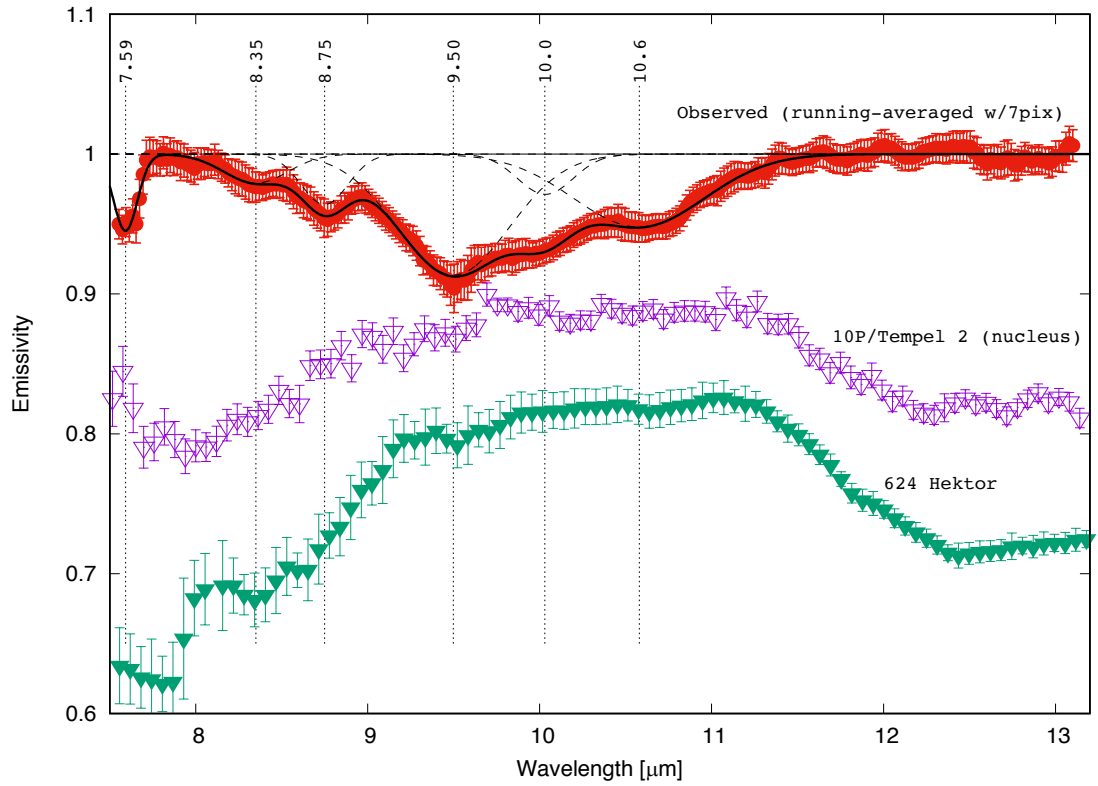

Figure 4: Normalized thermal emissivity spectrum of P/2016 BA₁₄ compared with the thermal emissivity spectra of comet 10P/Tempel 2 (nucleus) and the D-type asteroid 624 Hekor (Kelley et al. 2017; Emery et al. 2006). The spectra of 10P/Tempel 2 and 624 Hektor are shown with offsets for sake of readability. Solid line is a linear combination of six Gaussian components fitted to the observed spectrum (each component is plotted by dashed line). The vertical dotted lines indicate peak wavelengths of the components.

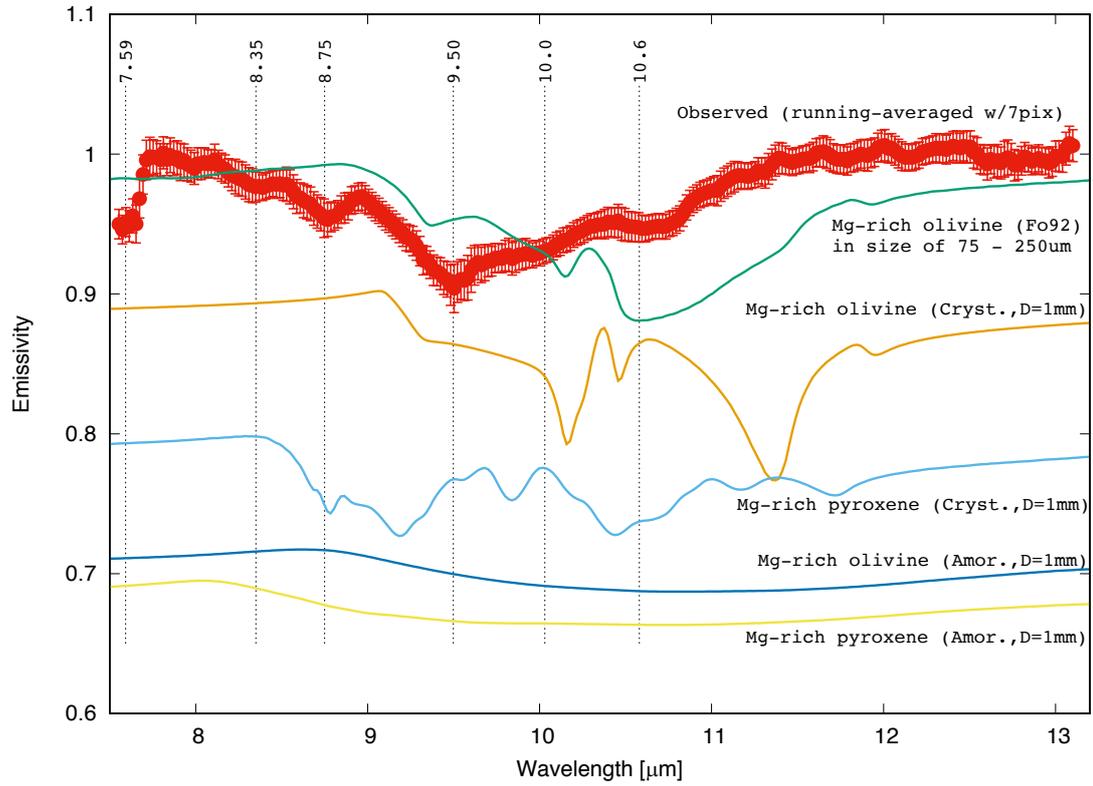

Figure 5: Normalized mid-infrared emissivity spectrum of P/2016 BA$_{14}$ compared with the estimated emissivity spectrum for a laboratory reflectance spectrum of coarse (75–250 μm-sized) grains of Mg-rich natural olivine (Fo92) and the modeled surface emissivity spectra of crystalline/amorphous Mg-rich silicate (olivine and pyroxene) grains. The modeling was based on the Hapke theory in combination with the Mie theory for 1 mm-sized porous grains.

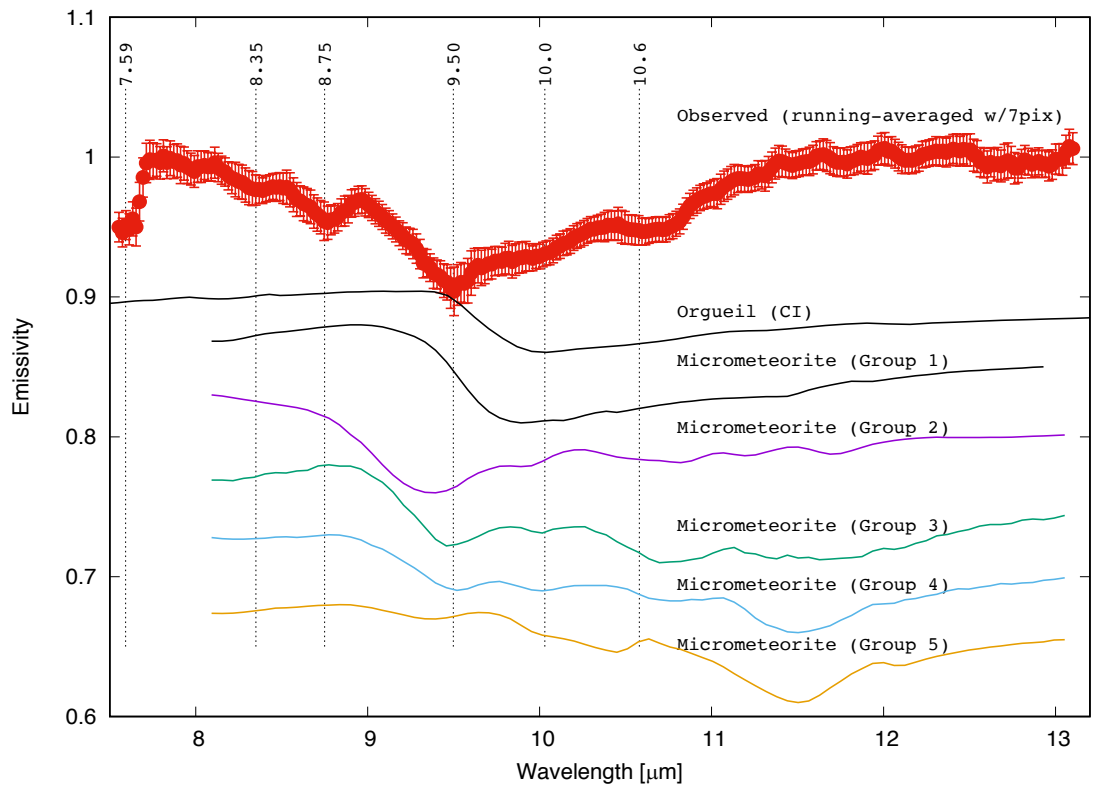

Figure 6: Normalized mid-infrared emissivity spectrum of P/2016 BA$_{14}$ compared with that of the chondritic meteorite Orgueil (CI) and the scaled $(1 - R)$ spectra of micrometeorites from Suttle et al. (2017), where $R$ denotes reflectance. All spectra other than that of P/2016 BA$_{14}$ are shifted for readability.

Acknowledgements

This paper is based on data collected by the Subaru Telescope, which is operated by the National Astronomical Observatory of Japan. We thank Fumihiko Usui for useful comments and discussion. This study is supported by Japan Society for the Promotion of Science (JSPS) KAKENHI Grant-in-Aid for Scientific Research (C) 17K05381, (B) 20H01943, (A) 19H00725 (T.O.), and (C) 20K14541 (Y.S.).

REFERENCES

Baldridge, A. M., Hook, S. J., Grove, C. I., & Rivera, G., 2009. The ASTER spectral library version 2.0. Remote Sensing of Environment, 113, 711–715.

Bardyn, A., Baklouti, D., Cottin, H., et al., 2017. Carbon-rich dust in comet 67P/Churyumov-Gerasimenko measured by COSIMA/Rosetta. Mon. Not. R. Astron. Soc. 469, S712–S722.

Bohren, C.F. & Huffman, D.R., 1983. Absorption and Scattering of Light by Small Particles. John Wiley & Sons, Inc.

Campins, H., de León, J., Licandro, J., et al., 2018. Compositional Diversity Among Primitive Asteroids. Primitive Meteorites and Asteroids, Physical, Chemical, and Spectroscopic Observations Paving the Way to Exploration (ed. Abreu, N.M.). Elsevier.

Che, C. & Glotch, T. D., 2012. The effect of high temperatures on the mid-to-far-infrared emission and near-infrared reflectance spectra of phyllosilicates and natural zeolites: Implications for martian exploration. Icarus, 218, 585–601.

Chihara, H., Koike, C., & Tsuchiyama, A., 2001. Low-Temperature Optical Properties of Silicate Particles in the Far-Infrared Region. Publ. Astron. Soc. Japan, 53, 243–250.

Cohen, M., Walker, R.G., Carter, B., et al., 1999. Spectral irradiance calibration in the infrared. X. a self-consistent radiometric all-sky network of absolutely calibrated stellar spectra. Astron. J. 117, 1864–1889.

Coates, J., 2000. Interpretation of Infrared Spectra: A Practical Approach. Encyclopedia of Analytical Chemistry (Meyers, R.A., Ed., John Wiley & Sons Ltd., Chichester), 10881–10882.

Dorschner, J., Begemann, B., Henning, T., Jäger, C., Mutschke, H., 1995. Steps toward interstellar silicate mineralogy. II. Study of Mg–Fe–silicate glasses of variable composition. Astron. Astrophys. 300, 503–520.

Draine, B.T., Li, A., 2007. Infrared emission from interstellar dust. IV. The silicategraphite-PAH model in the post-SPITZER era. Astrophys. J. 657, 810–837.

Emerich, C., Lamarre, J.M., Moroz, V.I., Combes, M., Sanko, N.F., Nikolsky, Y.V., Rocard, F., Gispert, R., Coron, N., Bibring, J.P., 1986. Temperature and size of the nucleus of Halley's comet deduced from IKS infrared VEGA 1 measurements. In ESLAB Symposium on the Exploration of Halley's Comet 250., December.

Emery, J.P., Cruikshank, D.P., Van Cleve, J., 2006. Thermal emission spectroscopy (5.2–38 μm) of three Trojan asteroids with the Spitzer Space Telescope: Detection of fine-grained silicates. Icarus 182, 496–512.

Evans, A., et al., 2005. Infrared spectroscopy of Nova Cassiopeiae 1993 –VI. A closer look at the dust. Mon. Not. R. Astron. Soc. 360, 1483–1492.

Fabian, D., Henning, T., Jäger, C., et al., 2001. Steps toward interstellar silicate mineralogy. VI. Dependence of crystalline olivine IR spectra on iron content and particle shape. Astron. Astrophys. 378, 228–238.

Farnham, T.L., Schleicher, D.G., Woodney, L.M., Birch, P. V., Eberhardy, C. A., Levy, L., 2001. Imaging and Photometry of Comet C/1999 S4 (LINEAR) Before Perihelion and After Breakup. Science 292, 1348–1353.

Fernandez, J.A. 2005. Comets. Astrophysics and Space Science Library. Springer Nature.

Gounelle, M. & Zolensky, M.E., 2014. The Orgueil meteorite: 150 years of history. Meteoritics & Planetary Science 49, 1769–1794.

Groussin, O., A'Hearn, M.F., Li, J.Y., Thomas, P.C., Sunshine, J.M., Lisse, C.M., Meech, K. J., Farnham, T.L., Feaga, L.M., Delamere, W.A., 2007. Surface temperature of the nucleus of Comet 9P/Tempel 1. Icarus 187, 16–25. https://doi.org/10.1016/j.icarus.2006.08.030.

Groussin, O., Sunshine, J.M., Feaga, L.M., Jorda, L., Thomas, P.C., Li, J.Y., A'Hearn, M.F., Belton, M.J.S., Besse, S., Carcich, B., Farnham, T.L., 2013. The temperature, thermal inertia, roughness and color of the nuclei of Comets 103P/Hartley 2 and 9P/Tempel 1. Icarus 222, 580–594. https://doi.org/10.1016/j.icarus.2012.10.003.

Hamilton, V.E., 2010. Thermal infrared (vibrational) spectroscopy of Mg–Fe olivines: A review and applications to determining the composition of planetary surfaces. Chemie der Erde 70, 7–33.

Hapke, B., 2012. Theory of Reflectance and Emittance Spectroscopy (2nd ed.), Cambridge Univ. Press. https://doi.org/10.1017/CBO9781139025683.

Hyland, M.G., Fitzsimmons, A., Snodgrass, C., 2019. Near-UV and optical spectroscopy of comets using the ISIS spectrograph on the WHT. MNRAS 484, 1347–1358.

Jäger, C., Molster, F. J., Dorschner, J., et al., 1998. Steps toward interstellar silicate mineralogy. IV. The crystalline revolution. Astron. Astrophys. 339, 904–916.

Kataza, H., Okamoto, Y., Takubo, S., et al., 2000. COMICS: the cooled mid-infrared camera and spectrometer for the Subaru telescope. Proc. SPIE 4008, 1144–1152.

Kelley, M.S.P., Woodward, C. E., Gehrz, R.D., Reach, W.T., Harker, D.E.,, 2017. Mid-infrared spectra of comet nuclei. Icarus 284, 344–358.

Lane, M.D. & Bishop, J.L., 2019. Mid-infrared (Thermal) Emission and Reflectance Spectroscopy. Remote Compositional Analysis (eds. Bishop, J.L., Bell III, J.F., Moersch, J.E.), Cambridge Univ. Press.

Levy, D. 1998. THE COLLISION OF COMET SHOEMAKER–LEVY 9 WITH JUPITER. Space Science Reviews 85, 523–545.

Li, J.-Y., et al., 2017. The Unusual Apparition of Comet 252P/2000 G1 (LINEAR) and Comparison with Comet P/2016 BA14 (PanSTARRS). Astrophys. J. 154, 136.

Lisse, C.M., et al., 2006. Spitzer spectral observations of the deep impact Ejecta. Science 313, 635–640.

Lisse, C.M., et al., 2007. Comparison of the composition of the Tempel 1 ejecta to the dust in Comet C/Hale–Bopp 1995 O1 and YSO HD 100546. Icarus 191, 223–240.

Mumma, M. J. & Charnley, S., 2011. The Chemical Composition of Comets – Emerging Taxonomies and Natal heritage. Annu. Rev. Astron. Astrophys., 49, 471–524.

Naidu, S.P., Benner, L.A.M., Brozovic, M., et al., 2016. High-resolution Goldstone radar imaging of comet P/2016 BA14 (Pan-STARRS). Americal Astron. Soc. DPS meeting #48, id.219.05.

Okamoto, Y.K., Kataza, H., Yamashita, T., et al., 2003. Improved performances and capabilities of the cooled mid-infrared camera and spectrometer (COMICS) for the Subaru telescope. Proc. SPIE 4841, 169–180.

Ootsubo, T., Kawakita, H., Shinnaka, Y., Watanabe, J., Honda, M., 2020, Unidentified infrared emission features in mid-infrared spectrum of comet 21P/Giacobini-Zinner. Icarus 338, 113450.

Paganini, L., Camarca, M.N., Mumma, M. J., Faggi, S., Lippi, M., Villanueva, G. L., 2019. Observations of Jupiter Family Comet 252P/LINEAR During a Close Approach to Earth Reveal Large Abundances of Methanol and Ethane. Astrophys. J. 158, 98.

Poch, O., Istiqomah, I., Quirico, E., et al. 2020. Ammonium salts are a reservoir of nitrogen on a cometary nucleus and possibly on some asteroids. Science, 367, eaaw7462.

Raponi, A., Ciarniello, M., Capaccioni, F., et al. 2020. Infrared detection of aliphatic organics on a cometary nucleus. Nat. Astron., 4, 500–505.

Schleicher, D.G., 2008. THE EXTREMELY ANOMALOUS MOLECULAR ABUNDANCES OF COMET 96P/MACHHOLZ 1 FROM NARROWBAND PHOTOMETRY. Astronomical J. 136, 2204–2213.

Sekanina, Z. & Krachat, R., 2014. DISINTEGRATION OF COMET C/2012 S1 (ISON) SHORTLY BEFORE PERIHELION: EVIDENCE FROM INDEPENDENT DATA SETS. Astro-ph (arXiv:1404.5968v6).

Suttle, M.D., Folco, L., Genge, M.J., Russell, S.S., 2020. Flying too close to the Sun – The viability of perihelion-induced aqueous alteration on periodic comets. Icarus 351, 113956.

Suttle, M.D., Genge, M.J., Folco, L., Russell, S.S., 2017. The thermal decomposition of fine-grained micrometeorites, observations from mid-IR spectroscopy. Geochimica et Cosmochimica Acta 206, 112–136.

Shinnaka, Y., Ootsubo, T., Kawakita, H., et al., 2018. Mid-infrared Spectroscopic Observations of Comet 17P/Holmes Immediately After Its Great Outburst in 2007 October. Astrophys. J. 156, 242.

Soderblom, L.A., Britt, D.T., Brown, R.H., Buratti, B.J., Kirk, R.L., Owen, T.C., Yelle, R.V., 2004. Short-wavelength infrared (1.3–2.6 μm) observations of the nucleus of Comet 19P/Borrelly. Icarus 167, 100–112. https://doi.org/10.1016/j.icarus.2003.08.019.

Tosi, F., Capaccioni, F., Capria, M.T., Mottola, S., Zinzi, A., Ciarniello, M., Filacchione, G., Hofstadter, M., Fonti, S., Formisano, M., Kappel, D., 2019. The changing temperature of the nucleus of comet 67P induced by morphological and seasonal effects. Nature Astronomy 453, 649–658. https://doi.org/10.1038/s41550-019-0740-0.

Vernazza, P., Delbo, M., King, P.L., et al., 2012. High surface porosity as the origin of emissivity features in asteroid spectra. Icarus 221, 1162–1172.